\documentclass[aps,nofootinbib,twocolumn,superscriptaddress]{revtex4}%
\usepackage{amssymb}
\usepackage{amsmath}
\usepackage{epsfig}
\usepackage{amsfonts}
\usepackage{graphicx}
\usepackage{hyperref}
\usepackage[dvipsnames]{xcolor}%
\providecommand{\U}[1]{\protect\rule{.1in}{.1in}}
\newtheorem{definition}{Definition}

\begin{document}
\title{Rotating Black Holes and Coriolis Effect}
\author{Xiaoning Wu}
\email{wuxn@amss.ac.cn}
\affiliation{Institute of Mathematics, Academy of Mathematics and System Science, CAS, China}
\author{Yi Yang}
\email{yiyang@mail.nctu.edu.tw}
\affiliation{Department of Electrophysics, National Chiao Tung University, Hsinchu, ROC}
\author{Pei-Hung Yuan}
\email{phyuan.py00g@nctu.edu.tw}
\affiliation{Institute of Physics, National Chiao Tung University, Hsinchu, ROC}
\author{Chia-Jui Chou}
\email{agoodmanjerry.ep02g@nctu.edu.tw}
\affiliation{Department of Electrophysics, National Chiao Tung University, Hsinchu, ROC}

\begin{abstract}
In this work, we consider the fluid/gravity correspondence for general
rotating black holes. By using the Petrov-like boundary condition in near
horizon limit, we study the correspondence between gravitational perturbation
and fluid equation. We find that the dual fluid equation for rotating black
holes contains a Coriolis force term, which is closely related to the angular
velocity of the black hole horizon. This can be seen as a dual effect for the
frame-dragging effect of rotating black hole under the holographic picture.

\end{abstract}
\maketitle



\noindent\textit{Introduction} It is well-known that AdS/CFT correspondence is
a great breakthrough on theoretical physics. This conjecture offers us a
powerful tool to study properties of strongly coupled systems. An important
application of this conjecture is the correspondence between gravity and fluid
dynamics. Such correspondence was first observed by Policastro, Son and
Starinets \cite{PSS01}. The main idea of the correspondence is that the
infrared behaviour of the dual theories should be governed by hydrodynamics.
Since the gravitational perturbation in bulk space-time could be identified
with the perturbation of the dual field theory on the time-like boundary,
there should be a natural relation between the long wavelength perturbed
Einstein equation and the hydrodynamical equation. By considering long
wavelength model of perturbation on black brane solutions, Son et. al.
established such correspondence and calculated the associated shear viscosity
of the dual fluid. During the last decade, this topic has attracted great
attention of researchers. Many interesting fluid phenomenon have been realized
holographically, e.g. turbulence \cite{Liu12} and Hall viscosity \cite{Hall}.
Since the gravity/fluid duality is a quite natural generalization of AdS/CFT
correspondence, it is reasonable to believe that such correspondence should
hold for general stationary black holes. Unfortunately, the original method in
\cite{PSS01} requires the long wavelength condition, so that it can only be
used to deal with black brane cases. Especially, the method can not be applied
to rotating black holes. In 2011, Strominger and his colleagues proposed a new
idea to realize the correspondence \cite{BKLS,BKLS11}. They found that, by
imposing a Petrov-like boundary condition in suitable circumstances, the
Einstein equation reduces to the Navier-Stokes equation in one lower
dimension. Mathematically, this method is much simpler than the original one
and can be generalized to the cases of more general black holes. The
fluid/gravity correspondence for general non-rotating black holes has been
established in \cite{Wu13}. In this paper, we consider the case of general
rotating black holes.

On the gravity side, rotating black holes have an important phenomenon, the
frame-dragging effect \cite{MTW}. Near horizon, the stationary observer will
be forced to rotate with the black hole because of the distorted space-time
geometry. In fact, such effect exists for any rotating massive objects and has
been observed by the GPB experiment \cite{GPB}. An interesting question is
what the dual effect for the frame-dragging on the field theory side is. In
this paper, we will study the gravity/fluid duality and focus on the physical
effect on the dual fluid caused by the rotation of black holes. We discover
that the dual fluid equation is an incompressible Navier-Stokes equation with
Coriolis force. Our result implies that the holographic dual effect of
frame-dragging is just the Coriolis force, at least at the hydrodynamical
limit level.

In this paper, we first consider some basic properties of background black
hole configurations. To describe a general stationary black hole, we use the
theory of isolated horizon which is developed by Ashtekar and other authors
\cite{AK04}. Then we consider the fluid/gravity correspondence for rotating
black holes by the Petrov-like boundary condition method. Finally, We conclude
our results.\newline\newline\noindent\textit{Asymptotic behaviour of metric
near horizon} In order to study the gravity/fluid correspondence in general
cases, one needs to consider the properties of general stationary black hole
configuration. To do this, Ashtekar's isolated horizon theory is a suitable
tool \cite{AK04}. This theory was developed by Ashtekar and other authors for
four dimensional at the end of last country. It can be shown that any
stationary black hole horizon satisfies the definition of isolated horizon.
Later, Lewandowski and his colleagues generalized it to general dimensions
\cite{KLP05}.

\begin{definition}
(Isolated Horizon in $(p+2)$-dimensional space-time)\newline Let $(M,g)$ be a
$(p+2)$-dim Einstein manifold with or without a cosmological constant.
$\mathcal{H}$ is a $(p+1)$-dim null hypersurface in $M$ {and} $l$ is the null
normal of $\mathcal{H}$. $\mathcal{H}$ is called an isolated horizon in $M$ if
\newline(1). there exists an embedding : $S\times[0,1]\to M$, $\mathcal{H}$ is
the image of this map, $S$ is a p-dimensional connected manifold and for every
maximal null curve in $\mathcal{H}$ there exists $x\in S$ such that the curve
is the image of $x\times[0,1]$; \newline(2). the expansion of $l$ vanishes
everywhere on $\mathcal{H}$; \newline(3). $R_{ab}l^{a} l^{b}|_{\mathcal{H}}%
=0$; \newline(4). let $\mathcal{D}$ denote the induced connection on
$\mathcal{H}$, $[\mathcal{L}_{l}, \mathcal{D}]=0$ holds on $\mathcal{H}$.
\end{definition}

With the help of Killing equation, Rachardaury equation and Einstein equation,
it is easy to show that any stationary black hole horizon satisfies above
definition. In the rest part of this paper, we will focus on the gravitational
perturbation near isolated horizon. As has been discussed in \cite{Wu13}, one
can establish a Bondi-like coordinates $\{t,r,x^{i}\}$ ($i=1,2,\cdots,p$) in a
neighbourhood of isolated horizon. Further more, there is a special set of
null tetrad $\{l,n,E_{I}\}$ ($I=1,2,\cdots,p$), namely Bondi tetrad, which can
be expressed by the Bondi-like coordinates as:
\begin{align}
n  &  =\partial_{r},\nonumber\\
l  &  =\partial_{t}+U\partial_{r}+X^{i}\partial_{i},\nonumber\\
E_{I}  &  =W_{I}\partial_{r}+e_{I}^{i}\partial_{i},\qquad I,i=1,2,\cdots,p,
\label{tetrad0}%
\end{align}
where $(U,X^{i},W_{I},e_{I}^{i})$ are functions of $(t,r,x^{i})$. As discussed
in \cite{Fr81}, the Bondi gauge implies $\hat{U}=\hat{X}^{i}=\hat{W_{I}}=0$.
According to Ref. \cite{AK04}, all hatted quantities are the initial data of
isolated horizon $\mathcal{H}$. In Bondi coordinates, the general form of
metric in the neighbourhood of horizon should be
\begin{equation}
(g^{\mu\nu})=\left(
\begin{array}
[c]{ccc}%
0 & 1 & \vec{0}\\
1 & 2U+W_{I}W_{I} & X^{i}+W_{I}e_{I}^{i}\\
\vec{0} & X^{j}+W_{I}e_{I}^{j} & e_{I}^{i}e_{I}^{j}%
\end{array}
\right)  , \label{metric}%
\end{equation}
where the Cartan structure equations tell us that the behaviour of the unknown
functions in metric are
\begin{align}
U  &  =\hat{\epsilon}r+\frac{1}{2}(\hat{R}_{nlnl}+2|\hat{\pi}|^{2}%
)r^{2}+O(r^{3}),\\
W_{I}  &  =-\hat{\pi}_{I}r+\frac{1}{2}(\hat{R}_{nInl}+2\hat{\theta}%
_{IJ}^{\prime}\hat{\pi}_{J})r^{2}+O(r^{3}),\nonumber\\
X^{i}  &  =-\hat{\pi}^{i}r+\frac{1}{2}(\hat{R}_{nInl}\hat{e}_{I}^{i}%
+2\hat{\theta}_{IJ}^{\prime}\hat{\pi}_{I}\hat{e}_{J}^{i})r^{2}+O(r^{3}%
),\nonumber\\
e_{I}^{i}  &  =\hat{e}_{I}^{i}-\hat{\theta}_{IJ}^{\prime}\hat{e}_{J}%
^{i}r+\frac{1}{2}({\hat{R}}_{nInJ}\hat{e}_{J}^{i}+2\hat{\theta}_{IJ}^{\prime
}\hat{\theta}_{JK}^{\prime}\hat{e}_{K}^{i})r^{2}+O(r^{3}),\nonumber
\end{align}
where ${\hat{e}}_{I}^{i}$ is the tetrad on the section of horizon,
$\theta_{IJ}^{\prime}:=\langle E_{I},\nabla_{J}n\rangle$ and $\epsilon
:=\langle n,\nabla_{l}l\rangle$ with ${\hat{\epsilon}}$ being the surface
gravity of horizon, $\pi_{I}:=\langle E_{I},\nabla_{l}n\rangle$ is just the
rotational 1-form potential in \cite{AK04}, which is related to the angular
momentum of horizon. $\pi_{I}\neq0$ implies that the black hole is rotating.
In previous works, all black holes considered are non-rotating \footnote{When
we finish this paper, there appears another paper \cite{1511.08002} discussing
Kerr/fluid Duality.}. The main concern of this paper is to study the effect
induced by the non-zero $\pi_{I}$.\newline\newline\noindent\textit{Brown-York
tensor of time-like boundary near horizon} When one studies the fluid/gravity
correspondence by using Strominger's Petrov-like boundary condition method
\cite{BKLS,BKLS11}, the asymptotic behaviour of extrinsic curvature of the
time-like boundary is crucial. The first reason is that, based on the
gauge/gravity dictionary, the radius of time-like boundary is related to the
energy scale of the dual field theory on the boundary. So the radius
approaching the horizon implies the low energy limit in the dual field
theory\footnote{Such limit also has been used to consider other topics about
black hole which is related with AdS/CFT correspondence, such as Kerr/CFT
correspondence \cite{Kerrcft}.}. The second reason is that the Brown-York
tensor corresponds to the energy-momentum tensor of the dual field theory. It
is well known that the dynamical equation of fluid comes from the conservation
law of the energy-momentum tensor, so we need to know the asymptotic behaviour
of Brown-York tensor. Such behaviour can be obtained by direct calculation
based on the asymptotic results of metric in last section. We summarize our
approach as follows: introduce a rescaling parameter $\lambda$ and consider a
time-like boundary $r=r_{c}$ in the neighbourhood of the horizon, then define
a new temporal coordinate $\tau=2{\hat{\epsilon}}\lambda^{2}t$ and
$r_{c}=2{\hat{\epsilon}}\lambda^{2}$ as well. After taking $\lambda
\rightarrow0$ limitation, it turns out the near horizon and non-relativistic
limit in the meanwhile.
\begin{align}
K_{\tau}^{\tau}=  &  \frac{1}{2\lambda}+\beta\lambda+O(\lambda^{3}%
),\label{K}\\
K_{i}^{\tau}=  &  -\hat{\pi}_{i}\lambda+\psi_{i}\lambda^{3}+O(\lambda
^{5}),\nonumber\\
K_{j}^{i}=  &  \xi_{j}^{i}\lambda+O(\lambda^{3}),\nonumber\\
K=  &  \frac{1}{2\lambda}+(\beta+\xi)\lambda+O(\lambda^{3}),\nonumber
\end{align}
where
\begin{align}
\beta=  &  -\frac{1}{4}(7\hat{R}_{nlnl}-3|{\hat{\pi}}|^{2}),\label{beta}\\
\psi_{i}=  &  -\frac{1}{2}\nabla_{i}(\hat{R}_{nlnl}-|{\hat{\pi}}|^{2}%
)-4{\hat{\pi}}^{j}\nabla_{\lbrack i}{\hat{\pi}}_{j]}\nonumber\\
&  +2{\hat{\epsilon}\hat{g}}_{ij}(\hat{R}_{nInl}{\hat{e}}_{I}^{j}-\hat{\theta
}_{IJ}^{\prime}{\hat{\pi}}_{I}{\hat{e}}_{J}^{j})+\frac{1}{2}{\hat{\pi}}%
_{i}(\hat{R}_{nlnl}+3|{\hat{\pi}}|^{2}),\nonumber\\
\xi_{j}^{i}=  &  -2{\hat{g}}^{ik}{\tilde{\nabla}}_{(j}{\hat{\pi}}_{k)}%
+2{\hat{\pi}}^{i}{\hat{\pi}}_{j}+2{\hat{\epsilon}\hat{g}}_{jm}\hat{\theta
}_{IJ}^{\prime}{\hat{e}}_{I}^{i}{\hat{e}}_{J}^{m},\nonumber
\end{align}
and $\xi=\xi_{i}^{i}$ also ${\tilde{\nabla}}$ is induced derivative on the
section of horizon. The asymptotic behaviour of associated Brown-York tensor
are
\begin{align}
t_{\tau}^{\tau}  &  =\xi\lambda+O(\lambda^{3}),\label{BY}\\
t_{i}^{\tau}  &  ={\hat{\pi}}_{i}\ \lambda+O(\lambda^{3}),\nonumber\\
t_{\tau}^{i}  &  =O(\lambda^{3}),\nonumber\\
t_{j}^{i}  &  =\frac{1}{2\lambda}\delta_{j}^{i}+\left[  \left(  \beta
+\xi\right)  \delta_{j}^{i}-\xi_{j}^{i}\right]  \lambda+O(\lambda
^{3}),\nonumber\\
t  &  =\frac{p}{2\lambda}+\left[  p\left(  \beta+\xi\right)  \right]
\lambda+O(\lambda^{3}).\nonumber
\end{align}
\newline\noindent\textit{Petrov-like boundary condition and gravitational
perturbation} Following gravity/fluid correspondence, one needs to consider
the gravitational perturbation in bulk space-time. A basic requirement for
perturbation is to satisfy regular condition at horizon. In previous work,
people solve the perturbation equation concretely to insure the regularity.
For general isolated horizon, this method fails to work. One needs other
method to insure the regularity of perturbation. Thanks for the results on
initial-boundary value problem \cite{IBVP}, one can insure such regularity by
imposing suitable boundary condition. One of possible choice is Strominger's
Petrov-like boundary condition \cite{BKLS,BKLS11}. The Petrov-like boundary
condition requires the perturbed Weyl curvature satisfy following condition on
time-like boundary :
\begin{equation}
\hat{C}_{lilj}=0,
\end{equation}
where $l^{a}$ is the out-pointed null normal of the time-like boundary. As the
original paper by Strominger et. al., the perturbation is introduced in terms
of Brown-York tensor:
\begin{equation}
t_{b}^{a}=t_{b}^{a(B)}+\sum_{k}t_{b}^{a(k)}\lambda^{k},
\end{equation}
where $t_{b}^{a(B)}$ is the Brown-York tensor for back ground space-time,
$t_{b}^{a(k)}$ are gravitational perturbations and $\lambda\ll1$ is the
perturbation parameter. Under such rescaling, the perturbed Petrov-like
boundary condition implies that%
\begin{align}
t_{j}^{i(1)}=  &  -2{\hat{g}}^{ik}{\tilde{\nabla}}_{(j}(t^{\tau(1)}+{\hat{\pi
}})_{k)}+2t^{\tau i(1)}\left(  t_{j}^{\tau(1)}+{\hat{\pi}}_{j}\right)
\nonumber\label{Petrov1}\\
&  +\frac{t^{(1)}}{p}\delta_{j}^{i}+\xi_{j}^{i}-\tilde{R}_{j}^{i}%
+\frac{2(p^{2}-3)}{p^{2}(p+1)}\Lambda\delta_{j}^{i}.
\end{align}
Compare with the result of \cite{Wu13}, it is clear that this equation reduce
to the non-rotating result if ${\hat{\pi}}_{I}=0$.\newline\newline%
\noindent\textit{The dual Navier-Stokes equation} With preparation in previous
sections, we are able to study the holographic dual of gravitational
perturbation. The basic AdS/CFT dictionary tells us that the Brown-York tensor
corresponds to the energy-momentum tensor of dual field theory. On the field
theory side, the hydrodynamic limit of the conservation law of energy-momentum
tensor should give the fluid equation. On the gravity side, the conservation
equation of Brown-York tensor is just the Codazzi equation on the time-like
boundary. Thus the hydrodynamic limit indeed corresponds to the near horizon
limit. So what one has to consider the near horizon limit of the Codazzi
equation, ${\bar{D}}_{a}t_{b}^{a}=0$, where ${\bar{D}}$ is the induced
derivative on time-like boundary. Since the inner geometry on time-like
boundary is fixed, the perturbed Codazzi equation are
\begin{equation}
0={\bar{D}}_{a}t_{b}^{a(B)}+{\bar{D}}_{a}t_{b}^{a(1)}\lambda+O(\lambda^{2}).
\end{equation}
Considering the $\tau$ component of Codazzi equation, the first nontrivial
equation is in the $O(\lambda^{-1})$,
\begin{equation}
{\tilde{\nabla}}_{i}({\hat{g}}^{ij}t_{j}^{\tau(1)})=0. \label{NS1}%
\end{equation}
For $i$ components of Codazzi equation, under the near horizon limit, the
first nontrivial equation is in the $O(\lambda)$,
\begin{align}
0=  &  \partial_{\tau}t_{i}^{\tau(1)}-2t^{\tau j(1)}{\tilde{\nabla}}_{(i}%
\hat{\pi}_{j)}-2t^{\tau j(1)}\tilde{\nabla}_{[i}\hat{\pi}_{j]}-\tilde{\nabla
}_{j}(\xi_{i}^{j}-t_{i}^{j(1)})\nonumber\label{Codazzi}\\
&  +{\tilde{\nabla}}_{i}\left(  \beta+\xi-\frac{1}{4}(\hat{R}_{nlnl}-|\hat
{\pi}|^{2})\right)  .
\end{align}
Combined with Eq.(\ref{Petrov1}) and Eq.(\ref{NS1}) and used the concrete
expression of $\beta$ and $\xi$ in Eq.(\ref{beta}), this equation becomes%
\begin{align}
0=  &  \partial_{\tau}t_{i}^{\tau(1)}+\tilde{\nabla}_{i}\frac{t^{(1)}}%
{p}+2t^{\tau j(1)}\tilde{\nabla}_{j}t_{i}^{\tau(1)}-4t^{\tau j(1)}%
\tilde{\nabla}_{[i}\hat{\pi}_{j]}\nonumber\\
&  -{\tilde{\nabla}}^{2}(t_{i}^{\tau(1)}+{\hat{\pi}}_{i})-\tilde{R}_{i}%
^{j}(t_{j}^{\tau(1)}+{\hat{\pi}}_{j})-\tilde{\nabla}_{j}\tilde{R}_{i}%
^{j}\nonumber\\
&  -{\tilde{\nabla}}_{i}(2\hat{R}_{nlnl}+4\tilde{\nabla}_{j}\hat{\pi}%
^{j}-5|\hat{\pi}|^{2}),
\end{align}
where ${\tilde{R}}_{i}^{j}$ is the Ricci curvature of the section metric
$\hat{g}_{ij}$.

Now let's identify the geometric quantities with hydrodynamic quantities based
on gauge/gravity dictionary. Since $t_{b}^{a}$ corresponds to the
energy-momentum tensor in the dual field theory, it should be identified with
the fluid energy-momentum tensor under the hydrodynamic limit. Thus we take
the standard identification \cite{BKLS11},
\begin{equation}
t_{i}^{\tau(1)}\rightarrow\frac{1}{2}v_{i},\qquad\frac{t^{(1)}}{p}%
\rightarrow\frac{P}{2},
\end{equation}
where $P$ is the pressure and $v_{i}$ is the velocity in the dual fluid. With
this identification, Eq.(\ref{NS1}) tells us that\ the dual fluid is
incompressible, i.e. $\tilde{\nabla}_{i}v^{i}=0,$ and the fluid equation can
be finally written as,
\begin{align}
0= &  \partial_{\tau}v_{i}+\tilde{\nabla}_{i}P+v^{j}\tilde{\nabla}_{j}%
v_{i}-{\tilde{\nabla}}^{2}v_{i}-\tilde{R}_{i}^{j}v_{j}-4v^{j}\tilde{\nabla
}_{[i}\hat{\pi}_{j]}\nonumber\\
&  -2{\tilde{\nabla}}^{2}\hat{\pi}_{i}-2\tilde{\nabla}_{j}\tilde{R}_{i}%
^{j}-2\tilde{R}_{i}^{j}\hat{\pi}_{j}\nonumber\\
&  -2\tilde{\nabla}_{i}(2\hat{R}_{nlnl}+4\tilde{\nabla}_{k}\hat{\pi}%
^{k}-5|\hat{\pi}|^{2})\nonumber\\
&  -4v^{j}\tilde{\nabla}_{[i}\hat{\pi}_{j]},\label{NSf}%
\end{align}
which can be realized as the forced incompressible Navier-Stokes equations.
The first line in Eq.(\ref{NSf}) are standard terms of Navier-Stokes equation
in curved space-time. The second and third lines in Eq.(\ref{NSf}) are total
divergence of some quantities which is only dependent on back ground geometry
and can be realized as external forces. The only unusual term is the fourth
line in Eq.(\ref{NSf}). An interesting recognizing is that this term has the
form of Coriolis force. According to Eq.(\ref{BY}) and gauge/gravity
dictionary, the vector ${\hat{\pi}}_{a}$ is the velocity of the reference
frame, and $d{\hat{\pi}}_{a}$ is just the angular velocity. In order to see
this, we consider the Gauss equation. Under near horizon limit, the perturbed
Gauss equation gives
\begin{align}
t_{\tau}^{\tau(1)} &  =-2\hat{g}^{ij}t_{i}^{\tau(1)}t_{j}^{\tau(1)}-2\hat{\pi
}_{I}\hat{e}_{I}^{i}t_{i}^{\tau(1)}-\xi+\tilde{R}\nonumber\\
&  =-\frac{1}{2}|v+{\hat{\pi}}|^{2}+\left(  -\xi+{\tilde{R}}+\frac{1}{2}%
|{\hat{\pi}}|^{2}\right)  .
\end{align}
Based on the AdS/CFT dictionary, $t_{\tau}^{\tau}$ corresponds to the energy
density of the dual fluid. Obviously, the first term can be recognized as the
non-relativistic kinematic energy. This is agrees with that the Navier-Stokes
equation describes the non-relativistic dynamics of fluid.

If the horizon section metric ${\hat{g}}_{ij}$ in a 5-dimension space-time is
flat (based on characteristic initial value problem \cite{Fr81}, such solution
exists, at least locally.), we can identify that
\begin{equation}
\mathbf{\Omega}=\tilde{\nabla}\times\mathbf{\hat{\pi}}.
\end{equation}
Then Eq.(\ref{NSf}) becomes%
\begin{equation}
\partial_{\tau}\mathbf{v}+\mathbf{v}\cdot\tilde{\nabla}\mathbf{v}%
+\tilde{\nabla}P-\tilde{\nabla}^{2}\mathbf{v}+2\mathbf{\Omega}\times
\mathbf{v}+\mathbf{f}=0\label{NS4}%
\end{equation}
where the external force terms are:
\begin{equation}
\mathbf{f}=-2\tilde{\nabla}^{2}\mathbf{\hat{\pi}}-2{\tilde{\nabla}}(2\hat
{C}_{nlnl}+4{\tilde{\nabla}}\cdot\mathbf{\hat{\pi}}-5|\mathbf{\hat{\pi}}%
|^{2}).
\end{equation}
Eq.(\ref{NS4}) is the standard incompressible Navier-Stokes equation in a flat
4-dim space-time with Coriolis force which is induced by the reference frame.
From this correspondence one can see that ${\hat{\pi}}_{a}$ should be realized
as the velocity of the reference frame. We thus show that the fluid/gravity
correspondence can be established for general stationary black holes including
rotation. Eq.(\ref{NSf}) is Navier-Stokes equation in a non-inertial
frame.\newline\newline\noindent\textit{Conclusion} In this paper we studied
the fluid/gravity correspondence for a general rotating black hole. We
considered a rotating black hole with an isolated horizon, which is more
general than an usual stationary horizon since only the geometry inside the
horizon is required to be stationary in this case. Further calculation has
shown that the fluid/gravity correspondence will be fail if one give up the
isolated condition. We showed that the fluid/gravity correspondence can also
be established for rotating black holes, especially Kerr black hole can be
seen as a special case of our result. Further more, a more interesting result
is that the dual fluid equation on rotating horizon contains a Coriolis force
term. The associated angular velocity is determined by a rotational 1-form
which is closely related to the angular velocity of horizon. Near a rotating
black hole horizon there is a famous effect called frame-dragging \cite{MTW},
i.e. the stationary observer will be forced to rotating with the horizon. We
thus proposed that the Coriolis effect should be the holographic dual of the
frame-dragging effect in a rotating black hole.

\noindent\textit{Acknowledgments} This work is partially supported by the
National Science Council (NSC 101-2112-M-009-005) and National Center for
Theoretical Science, Taiwan. X. Wu is supported by the National Natural
Science Foundation of China (Grant Nos. 11075206 and 11175245).

\end{document}